\documentclass[
aps,
twocolumn,
superscriptaddress,
prl,
amsmath,amssymb
]{revtex4-1}

\usepackage{graphicx}
\usepackage{dcolumn}
\usepackage{bm}
\usepackage{color}
\usepackage{hyperref}

\begin{document}

\title{Optically Levitated Nanodumbbell Torsion Balance and GHz Nanomechanical Rotor}

\author{Jonghoon Ahn}
 \affiliation{School of Electrical and Computer Engineering, Purdue University, West Lafayette, IN 47907, USA}

\author{Zhujing Xu}
 \affiliation{Department of Physics and Astronomy, Purdue University, West Lafayette, IN 47907, USA}

\author{Jaehoon Bang}
 \affiliation{School of Electrical and Computer Engineering, Purdue University, West Lafayette, IN 47907, USA}

\author{Yu-Hao Deng}
 \affiliation{State Key Lab for Mesoscopic Physics and School of Physics, Peking University, Beijing 100871, China}

\author{Thai M. Hoang}
 \thanks{Current address: Sandia National Laboratories, Albuquerque, NM 87123}
 \affiliation{Department of Physics and Astronomy, Purdue University, West Lafayette, IN 47907, USA}

\author{Qinkai Han}
\email{hanqinkai@mail.tsinghua.edu.cn}
 \affiliation{The State Key Laboratory of Tribology, Tsinghua University, Beijing 100084, China}

\author{Ren-Min Ma}
\email{renminma@pku.edu.cn}
 \affiliation{State Key Lab for Mesoscopic Physics and School of Physics, Peking University, Beijing 100871, China}
  \affiliation{Collaborative Innovation Center of Quantum Matter, Beijing 100871, China}

\author{Tongcang Li}
 \email{tcli@purdue.edu}
  \affiliation{School of Electrical and Computer Engineering, Purdue University, West Lafayette, IN 47907, USA}
  \affiliation{Department of Physics and Astronomy, Purdue University, West Lafayette, IN 47907, USA}
 \affiliation{Purdue Quantum Center, Purdue University, West Lafayette, IN 47907, USA}
 \affiliation{Birck Nanotechnology Center, Purdue University, West Lafayette, IN 47907, USA}

\date{\today}

\begin{abstract}
Levitated optomechanics has great potentials in precision measurements, thermodynamics, macroscopic quantum mechanics and quantum sensing. Here we synthesize and optically levitate silica nanodumbbells in high vacuum.
With a linearly polarized laser, we observe the torsional vibration of an optically levitated nanodumbbell in vacuum.  The linearly-polarized optical tweezer provides  a restoring torque to confine the orientation of the nanodumbbell, in analog to the torsion wire which provides restoring torque for suspended lead spheres in the Cavendish torsion balance. Our calculation shows its  torque detection sensitivity can exceed that of the current state-of-the-art  torsion balance by several orders. The levitated nanodumbbell torsion balance provides rare opportunities to observe the Casimir torque and probe the  quantum nature of gravity as proposed recently. With a circularly-polarized laser, we drive  a 170-nm-diameter nanodumbbell  to rotate beyond 1~GHz, which is the fastest nanomechanical rotor realized to date. Our calculations show that smaller silica nanodumbbells can sustain rotation frequency beyond 10 GHz. Such ultrafast rotation may be used to study material properties and probe vacuum friction.
\end{abstract}
\maketitle

Levitated optomechanical systems  provide a powerful platform for precision measurements  with great isolation from the thermal environment \cite{yin2013review,ashkin1976optical,Li2011,Romero10,Chang2010Cavity,frimmer2017controlling}.
Optically levitated nano- and microspheres have been used to demonstrate force  sensing at the level of $10^{-21}$ N \cite{ranjit2016zeptonewton} and to search for interactions associated with dark energy \cite{rider2016search}.
Optically trapped nanoparticles can also be driven to rotate at high speed. Previously,  a rotation frequency of about 10~MHz has been achieved \cite{Arita2013laser,kuhn2017optically,monteiro2018optical}. It is desirable to increase the   rotation frequency further for studying material properties under extreme conditions \cite{schuck2018ultrafast,nagornykh2017optical} and probing vacuum friction \cite{Zhao2012rotational,Guo2014giant}.

\begin{figure}[htb]
	\includegraphics[scale=0.45]{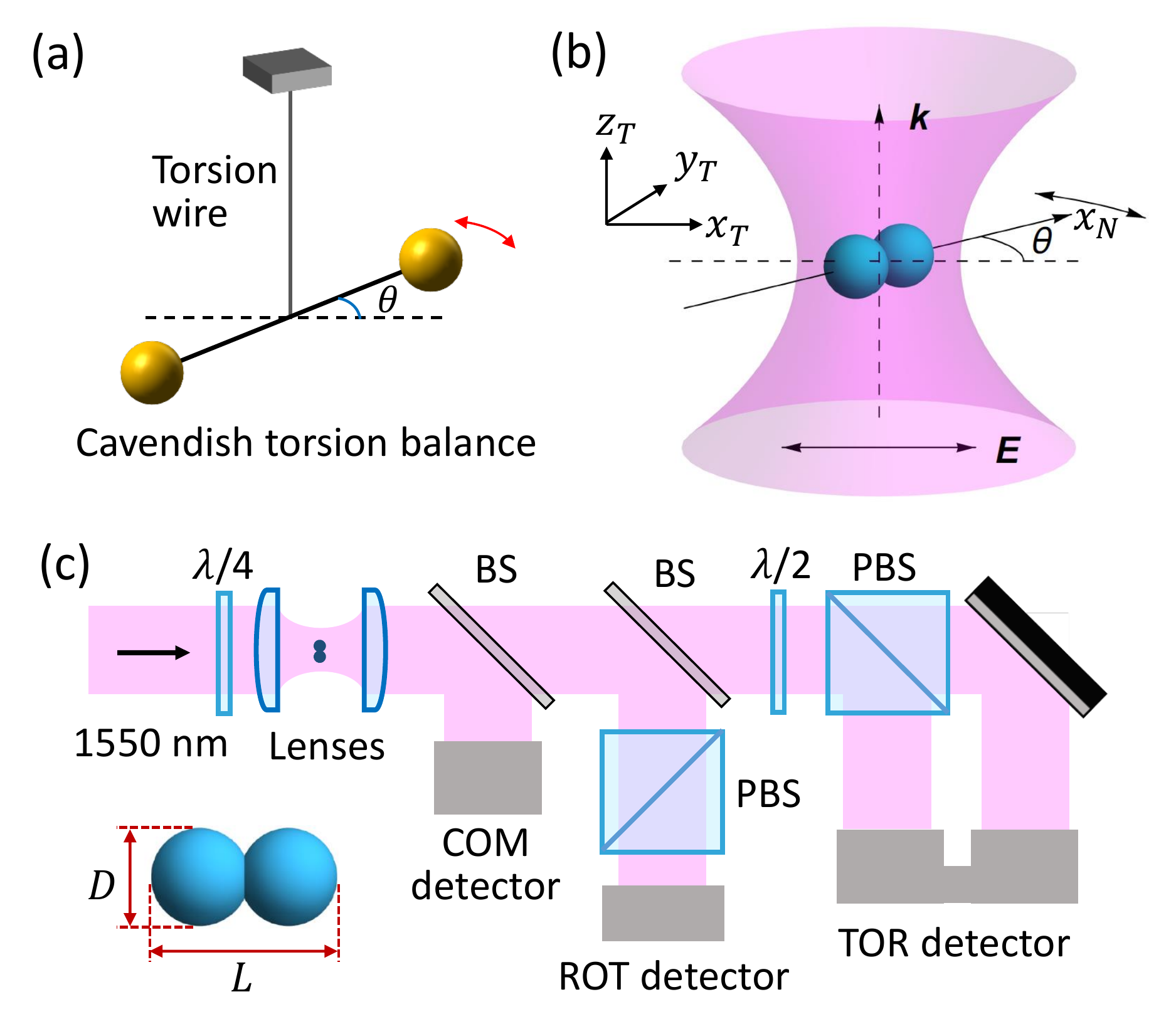}
	\caption{(a) A simplified diagram of the original Cavendish torsion balance that has two lead spheres suspended by a copper silvered torsion wire. (b) A  nanodumbbell levitated by a linearly polarized optical tweezer in vacuum. The linearly polarized optical tweezer provides the restoring torque that confines the orientation of the nanodumbbell.  $x_T$, $y_T$, $z_T$ are Cartesian coordinates of the trapping laser. $x_T$ is parallel to the electric field $\bf E$ of the incoming linearly polarized laser, and $z_T$ is parallel to the wave vector $\bf k$ of the laser. $x_N$ is parallel to the long axis of the nanodumbbell. The angle between $x_T$ and $x_N$ is $\theta$. (c) A simplified diagram for detecting the center-of-mass (COM) motion, the torsional (TOR) vibration, and the rotation (ROT)  of a levitated nanodumbbell. The nanodumbbell is trapped at the focus of the lenses. The laser beam is initially linearly polarized. A quarter-wave ($\lambda/4$) plate is used to control its polarization. BS: non-polarizing beam splitter; PBS: polarizing beam splitter; $\lambda/2$: half-wave plate. Inset: A nanodumbbell with  diameter  $D$ and  length $L$ created by attaching two identical nanospheres.
}	
\label{fig:scheme1}
\end{figure}

Recently, a novel ultrasensitive  torsion balance with an optically levitated nonspherical nanoparticle was proposed  \cite{Hoang2016torsional}, utilizing the coupling between the spin angular momentum of photons and the mechanical motion of the nanoparticle \cite{Hoang2016torsional,shi2013coupling,lechner2013cavity,Stickler2016Rotranslational}.
Torsion balances have enabled many breakthroughs in the history of  modern physics \cite{cavendish1798experiments,eotvos1890,beth1936mechanical,adelberger2009torsion}. For example, the Cavendish torsion balance (Fig. \ref{fig:scheme1}(a))   determined  the gravitational constant and the density of the Earth \cite{cavendish1798experiments}.
An optically levitated nanoscale torsion balance can provide a rare opportunity to detect the Casimir torque  \cite{Xu2017detecting,Parsegian1972dielectric,Barash1978moment,Enk1995Casimir}, and test the quantum nature of gravity \cite{Carlesso2018when,Marletto2017Gravitationally,Bose2017spin}. An essential step towards these goals is to optically trap  a well-defined nonspherical nanoparticle in high vacuum. However, optically trapped nonspherical nanoparticles such as nanodiamonds  and silicon nanorods in former experiments were lost at about 1 torr due to laser heating \cite{Neukirch2015,hoang2015observation,Hoang2016torsional,Rahman2016burning,kuhn2017optically}. Additionally, several years ago, levitated  nanodumbbells were theoretically proposed to study many-body phase transitions \cite{lechner2013cavity}. To our best knowledge, however, there was no report on optical levitation of a nanodumbbell in vacuum prior to this work.

In this letter, we synthesize silica nanodumbbells with two different methods and optically trap them in high vacuum.
  With a linearly polarized laser (Fig. \ref{fig:scheme1}(b)), we observe the torsional vibration of a levitated nanodumbbell in high vacuum, which is an important step towards probing the Casimir torque \cite{Xu2017detecting,Parsegian1972dielectric,Barash1978moment,Enk1995Casimir} and the quantum nature of gravity \cite{Carlesso2018when,Marletto2017Gravitationally,Bose2017spin}.  With a circularly polarized laser, we drive the nanodumbbell to rotate beyond 1 GHz, which is the highest mechanical rotation frequency reported to date.

In a linearly polarized optical tweezer, the long axis of a nanodumbbell will tend to align with the polarization direction of the trapping laser (Fig. \ref{fig:scheme1}(b)). This is because the polarizability of the nanodumbbell along its long axis is larger than the polarizability perpendicular to its long axis. If the nanodumbbell is not aligned with the polarization direction of the optical tweezer, it will twist the polarization of the optical tweezer (Fig. \ref{fig:scheme1}(b)), as an analog of twisting  the torsion wire by the lead spheres in the original Cavendish torsion balance (Fig. \ref{fig:scheme1}(a)). If the optical tweezer is circularly polarized, the nanodumbbell will be driven to rotate at high speed. The torsional vibration or rotation of the nanodumbbell can be detected by monitoring the change of the polarization of the trapping laser (Fig. \ref{fig:scheme1}(c)).

\begin{figure}
	\includegraphics[scale=0.55]{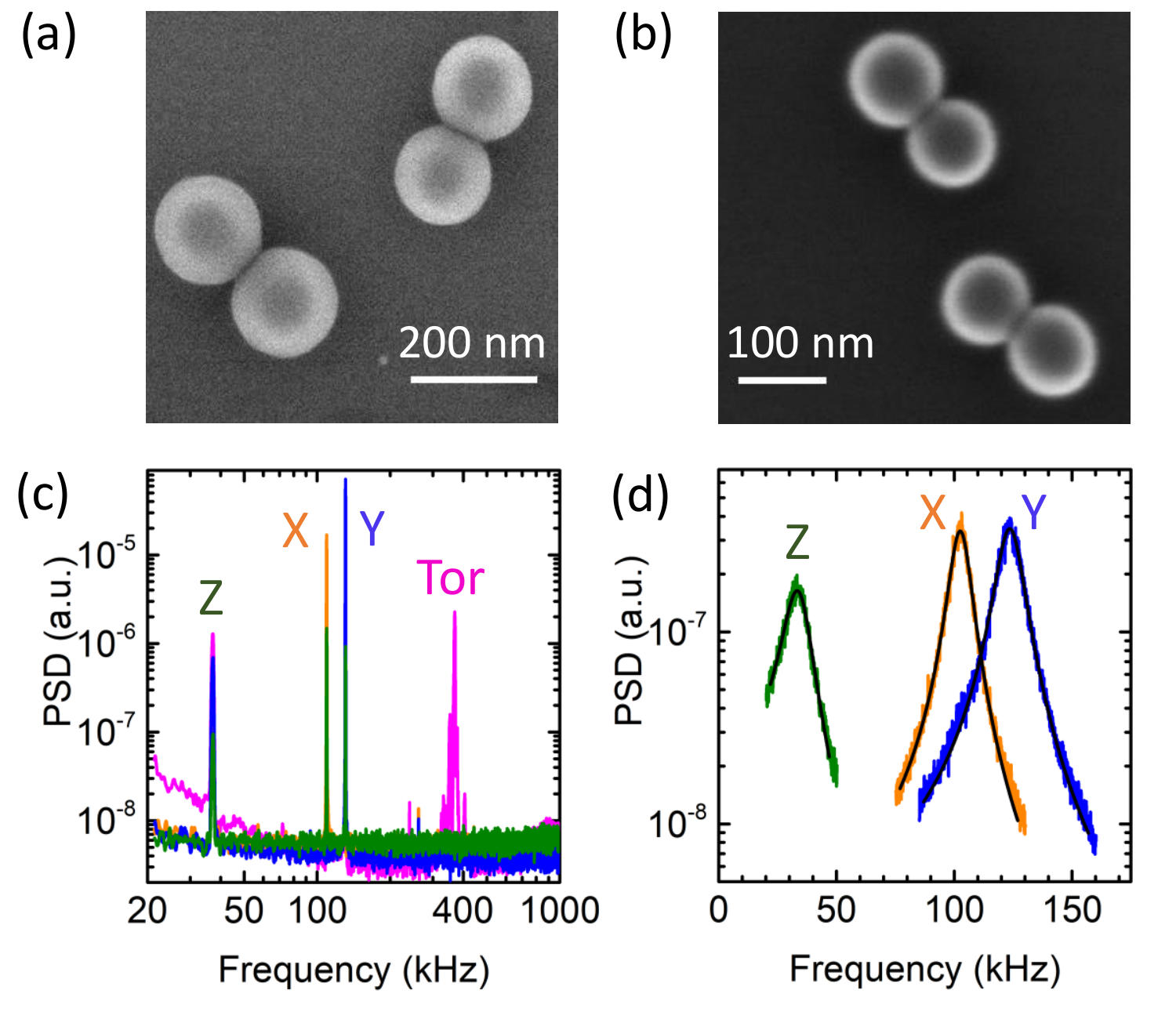}
	\caption{(a), (b): SEM images of silica nanodumbbells with two different sizes. The scale bar is 200~nm in (a), and 100~nm in (b).  (c) Measured power spectrum densities (PSD)  of the torsional vibration (labeled by `Tor') and the translational vibrations (labeled by `X', `Y', `Z') of a 170 nm-diameter   nanodumbbell optically levitated at $5 \times 10^{-4}$ torr. (d) Measured PSD of the translational vibrations of the nanodumbbell levitated at 10 torr. The black curves are Lorentz fits.
	}	
	\label{Fig:exptorsional}
\end{figure}

We have synthesized pure silica nanodumbbells using chemical and physical methods. To synthesize nanodumbbells with diameter $D$ and length $L$ (inset of Fig. \ref{fig:scheme1}(c)), we first synthesize silica  cores with a diameter of $d=L-D$. For example, we add 1 mL
tetraethyl orthosilicate (TEOS) to a  mixture of ammonia (4.86 mL), pure water (2.98 mL) and ethanol (100 mL) under stirring for 48 h to synthesize $d=80$~nm nanospheres. Then 10 mL acetone is added into the  solution and stirred for 24 h to induce aggregation. Next, a small amount of TEOS is added under stirring for another 24 h to grow the silica shells \cite{Johnson2005synthesis}. The precipitate
of silica dumbbells was purified by washing with ethanol in an autoclave at 90 $^\circ$C
and centrifugation.
 This chemical method can synthesize a large quantity  of silica nanodumbbells \cite{Johnson2005synthesis}, but is demanding.  So we also develop a physical  method to assemble nanodumbbells.  In this method, we first prepare a colloidal suspension of   silica nanospheres in water. We then generate water microdroplets in air with an ultrasonic nebulizer \cite{hoang2015observation}. By controlling the concentration of silica nanospheres, a fraction of  water microdroplets ($\sim 5 \mu$m in diameter) contain 2 silica nanospheres in them.  Two   nanospheres in the same microdroplet will form a nanodumbbell as the  water evaporates.
Fig. \ref{Fig:exptorsional}(a), \ref{Fig:exptorsional}(b) show SEM images of our nanodumbbells with two different sizes. Their aspect ratio ($L/D$) is between 1.9 and 2.

\begin{figure}[btp]
\setlength{\unitlength}{1cm}
\includegraphics[scale=0.46]{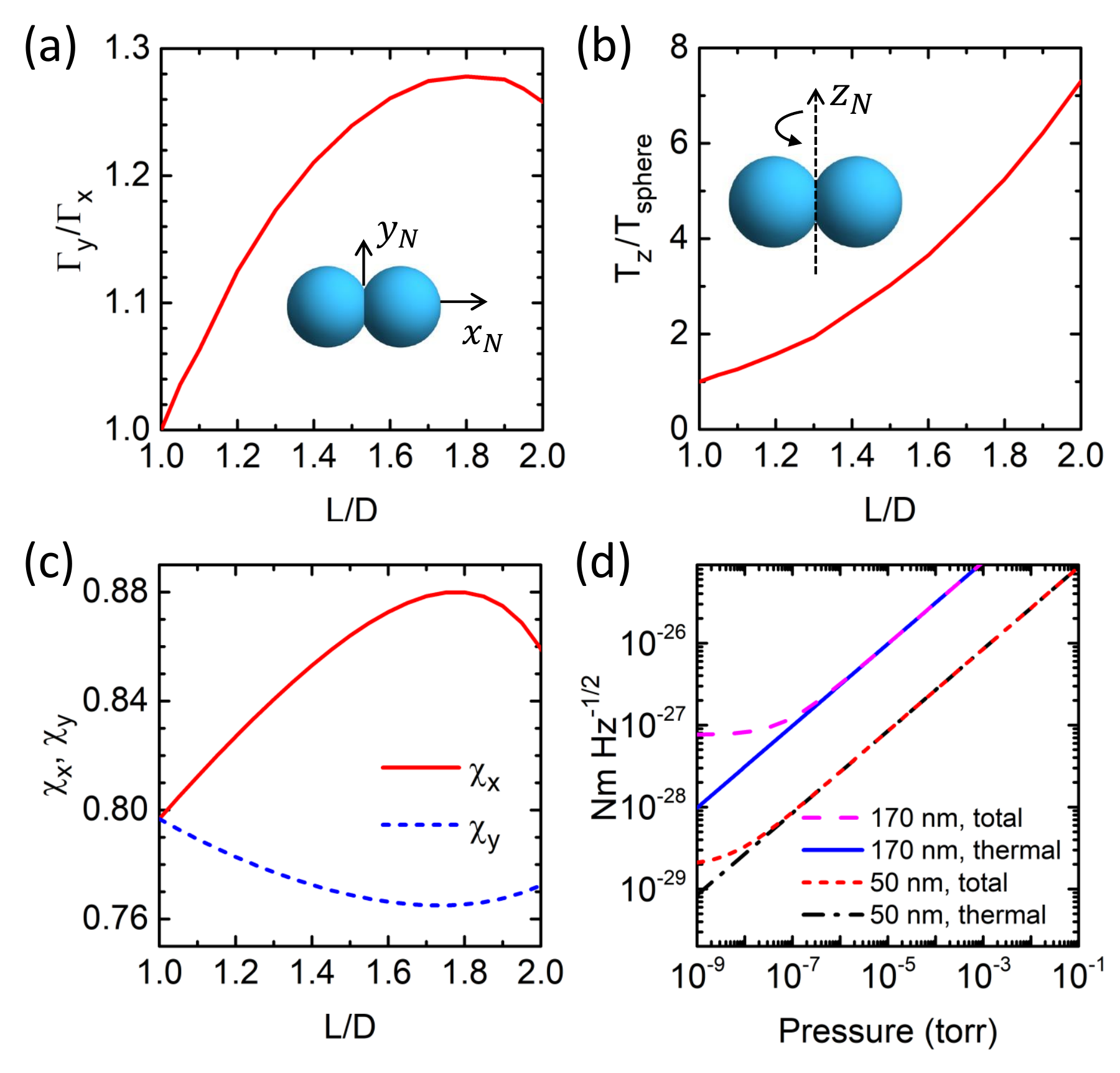}
\caption{ (a) The ratio of air damping coefficients for translational motions perpendicular or parallel to its long axis ($x_N$-axis) as a function of the aspect ratio ($L/D$) of a nanodumbbell.  (b) Calculated normalized drag torque of the rotation of a levitated nanodumbbell around $z_N$ axis as a function of the aspect ratio. (c) Effective susceptibilities of a silica nanodumbbell parallel ($\chi_x$) or perpendicular ($\chi_y$) to its long axis. (d) Calculated torque detection sensitivity of a levitated nanodumbbell with $D=170$ nm or $D=50$ nm as a function of air pressure. We assume $L/D=1.9$ in the calculations. The optical tweezer is assumed to be a focused 500 mW, 1550 nm laser with a waist of 820 nm.}
\label{fig:theory}
\end{figure}

To optically levitate a silica nanodumbbell in vacuum, a 500~mW, 1550~nm laser  is tightly focused with an NA = 0.85 objective lens in a vacuum chamber.  The laser is initially linearly-polarized , and its polarization can be controlled with a quarter wave plate (Fig. \ref{fig:scheme1}(c)). Silica nanodumbbells are delivered into the optical trap  at atmospheric pressure with an ultrasonic nebulizer \cite{hoang2015observation}. Once a nanoparticle is trapped, we evacuate the vacuum chamber to below 0.01 torr, and then increase the  pressure back to desired levels for measurements. This procedure removes extra nanoparticles in the chamber.
To monitor the trapping process, a 532 nm laser is applied on the nanoparticle and the scattered light is viewed using a  camera. We verify the trapped nanoparticle is a nanodumbbell by checking the  ratios of damping coefficients for translational motions along different directions. The motion of the levitated nanoparticle changes the direction and polarization of the laser beam, which allows us to monitor the motion of a nanodumbbell with the same  1550~nm trapping laser (Fig. \ref{fig:scheme1}(c)) \cite{Hoang2016torsional}. Fig. \ref{Fig:exptorsional}(c) shows the  power spectrum density (PSD) of both  torsional (TOR) vibration and center-of-mass (COM) vibration of a levitated 170-nm-diameter nanodumbbell in vacuum at $5 \times 10^{-4}$ torr. This is an important step towards quantum ground state cooling of the torsional vibration and testing recent theoretical proposals \cite{Xu2017detecting,Carlesso2018when,Marletto2017Gravitationally,Bose2017spin}.

\begin{figure*}[tb]
\includegraphics[scale=0.5]{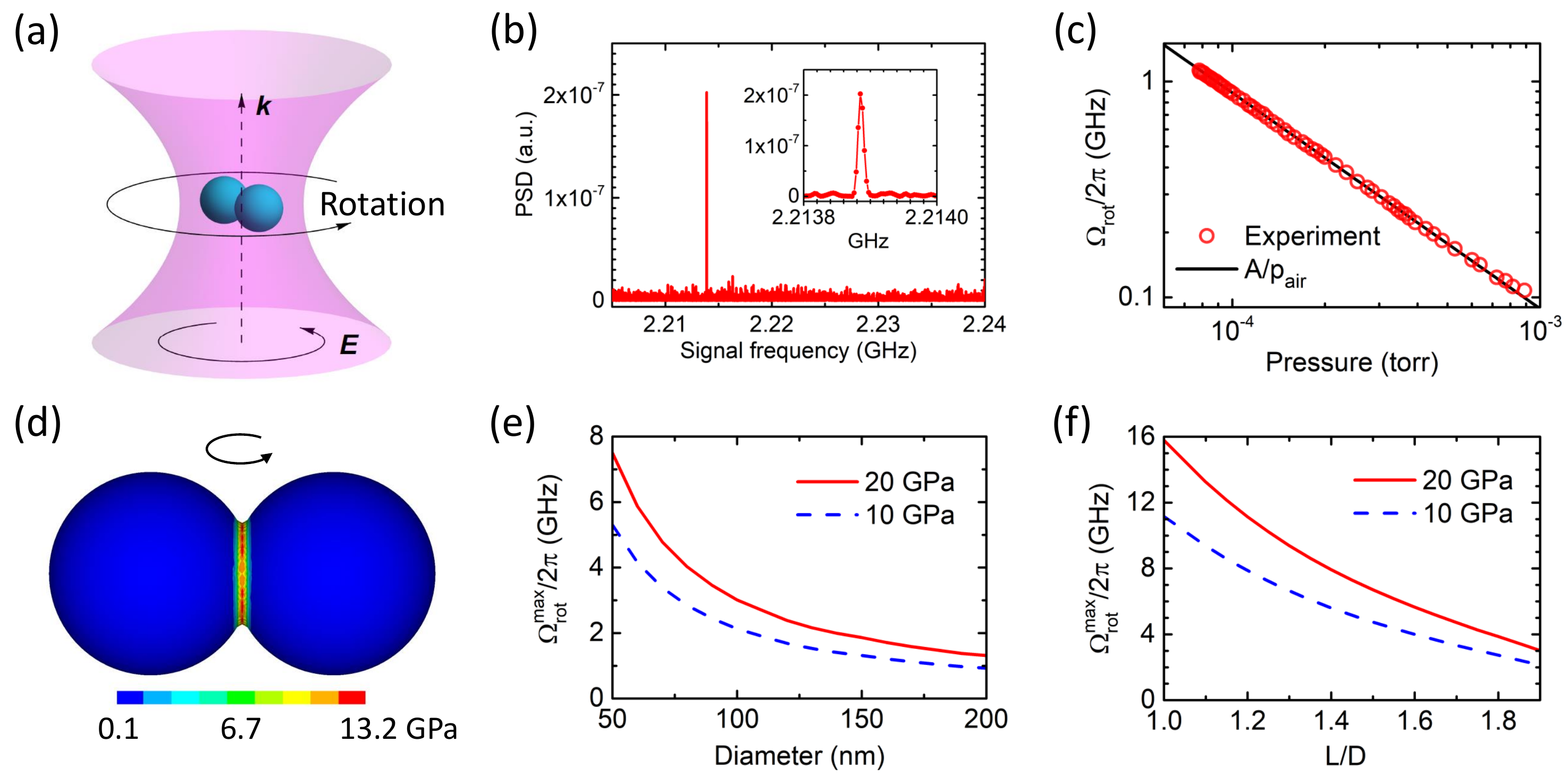}
\caption{(a) A nanodumbbell levitated by a circularly polarized laser will rotate. (b) Measured PSD of the rotation signal of a nanodumbbell. It has a sharp peak near 2.2 GHz. (c) Measured rotation frequency of the nanodumbbell as a function of pressure. (d) Calculated stress distribution of  a nanodumbbell ($D= 170$ nm, $L/D=1.9$) rotating at 1.2 GHz. (e) Calculated ultimate rotation frequency  as a function of the diameter of the nanodumbbell when $L/D=1.9$. (f) Calculated ultimate rotation frequency  as a function of the aspect ratio of the nanodumbbell when $D=100$ nm. The ultimate tensile strength is assumed be 20 GPa (red solid line) or 10 GPa (blue dashed line).}
\label{fig:rotation}
\end{figure*}

A levitated nanodumbbell will have anisotropic damping rates for translational motions in air if its orientation is fixed. We use the direct simulation Monte Carlo (DSMC) method to obtain the damping force and damping torque of a nanodumbbell  in the free molecular flow regime \cite{Mackowski2004montecarlo,Chan1981free,Corson2017calculating}.
In the simulation, molecules with uniform spatial distribution are launched from a spherical surface enclosing the nanodumbbell. The speeds of these molecules satisfy a shifted Maxwell distribution to include the effect of the motion of the nanodumbbell \cite{Corson2017calculating}.
Fig. \ref{fig:theory}(a) shows the calculated ratio of  damping rates of a nanodumbbell moving along ($\Gamma_x$) and  perpendicular ($\Gamma_y$) to its axial direction. The calculated ratio is $\Gamma_y/\Gamma_x=\Gamma_z/\Gamma_x=1.276$ when $L/D=1.9$, and $\Gamma_y/\Gamma_x=\Gamma_z/\Gamma_x=1.258$ when $L/D=2$. The measured ratios for the data of a 170 nm nanodumbbell shown in Fig. \ref{Fig:exptorsional}(d)  are $\Gamma_y/\Gamma_x=1.25 \pm 0.01$ and $\Gamma_z/\Gamma_x=1.27 \pm 0.02$, which agree excellently with our theoretical predictions.  The DSMC method is also utilized to obtain the drag torque $T_z$ on a nanodumbbell rotating at speed $\Omega$. We then calculate the ratio $T_z/T_{sphere}$, where $T_{sphere}=\pi \mu D^4 \Omega/(11.976 \lambda_{M})$ is the drag torque on a single sphere with diameter $D$ rotating at the same speed $\Omega$ \cite{Corson2017calculating}. Here $\mu$ is the viscosity of air, and $\lambda_{M}$ is the mean free path of air molecules which is inversely proportional to air pressure $p_{air}$. The calculated results are shown in Fig. \ref{fig:theory}(b). The damping rate of the rotation or torsional vibration  around the $z$ axis is $\Gamma_{\theta}=T_z/(I_z \Omega)$, where $I_z$ is its moment of inertia.

When the size of a silica nanodumbbell is much smaller than the wavelength of the trapping laser, the dipole approximation can be applied. The complex amplitude of the induced dipole of the nanodumbbell is $\textbf{p}= \alpha_x E_x \hat{x}_N + \alpha_y E_y \hat{y}_N+\alpha_z E_z \hat{z}_N$, where the complex amplitude of the electric field of the laser beam $\textbf{E}$ is decomposed into components along the principle axes of the nanodumbbell. $\hat{x}_N$ is in the direction along the long axis of the nanodumbbell. The components of the optical force $F_j$ and the optical torque $M_j$  acting on the nanodumbbell can be expressed as \cite{Trojek2012Optical}: $F_j =\frac{1}{2} Re \{\textbf{p}^* \cdot \partial_j \textbf{E}\}$ and
$M_j =\frac{1}{2} Re \{\textbf{p}^* \times \textbf{E}\}_j$.
The quasi-static polarizability  $\alpha^0_j$ ($j=x, y, z$) of a nanodumbbell   can be calculated assuming the electric field is static \cite{Pitkonen2006}.    Fig. \ref{fig:theory}(c) shows the effective susceptibilities  $\chi_x=\alpha^0_x/(\epsilon_0 V)$, $\chi_y=\alpha^0_y/(\epsilon_0 V)$  of the nanodumbbell as a function of the aspect ratio. Here  $\epsilon_0$ is the permittivity of vacuum, and $V$ is the volume of the nanodumbbell. $(\chi_x-\chi_y)/\chi_y=0.14$ when $L/D=1.9$.
The dipole approximation can be improved by including the effects of radiation reaction due to the  oscillation of the electric field in a laser beam. Then the polarizability  is  \cite{Draine1988,Trojek2012Optical,Dholakia2010}:
$\alpha_j=\alpha^0_j/ [1-i k^3_0 \alpha^0_j/(6\pi \epsilon_0)]$, where $k_0$ is the wave number. The real part of the polarizability  $Re[\alpha_j] \approx \alpha^0_j$ is responsible for optical confinement and alignment, while  the imaginary part $Im[\alpha_j]$ is important for optically rotating a nanodumbbell.

With calculated  damping rates and polarizabilities, we can calculate the torque detection sensitivity of a levitated nanodumbbell. The  minimum torque that it can measure limited by thermal noise is \cite{haiberger2007highly}: $M_{th}=\sqrt{4k_B T_{env} I_z \Gamma_{\theta}/\Delta t}$, where $T_{env}$ is the environmental temperature. In ultrahigh vacuum, the thermal noise from the residual air molecules becomes negligible and the minimum torque it can detect will be limited by the shot noise of the laser beam \cite{Xu2017detecting}: $M_{rad}=(\chi_x-\chi_y) k^2_0 V \hbar \sqrt{J_p/(3\pi \Delta t) }$. Here,  $J_p=I_{laser}/(\hbar \omega_L)$ is the photon flux. $I_{laser}$ is the laser intensity, and $\omega_L$ is the angular frequency of the laser. As shown in Fig. \ref{fig:theory}(d), a nanodumbbell with
$D=170$ nm and $D=50$ nm trapped in a 500 mW laser will have
a torque detection sensitivity  about $10^{-27} {\rm Nm/\sqrt{Hz}}$ and $10^{-29} {\rm Nm/\sqrt{Hz}}$, respectively. Remarkably, a levitated nanodumbbell at $10^{-4}$ torr is already much more sensitive than the state-of-the-art nanofabricated torsion balance, which has achieved a torque sensitivity of $10^{-22} {\rm N}{\rm m}/\sqrt{\rm Hz}$ at room temperature, and $10^{-24} {\rm N} {\rm m}/\sqrt{\rm Hz}$ at 25 mK in a dilution refrigerator \cite{kim2016approaching}.

While a nanodumbbell levitated by a linearly-polarized optical tweezer can be an ultrasensitive  nanoscale torsion balance, it will become an ultrafast nanomechanical rotor in a circularly-polarized optical tweezer (Fig. \ref{fig:rotation}(a)). The frequency of the detected signal will be twice  the rotation frequency of the nanodumbbell due to the symmetry of its shape. Fig. \ref{fig:rotation}(b) shows a PSD of the
rotation of a 170 nm-diameter nanodumbbell at $7.9\times10^{-5}$ torr. The detected signal has a sharp peak near 2.2 GHz. This shows the nanodumbbell rotates at 1.1 GHz, which is much faster than that from former experiments \cite{Arita2013laser,kuhn2017optically,monteiro2018optical}. Remarkably, the rotation is also very stable. The measured linewidth of the signal is less than 20 kHz,  limited by the resolution of our spectrum analyzer, which leads to a rotating quality factor of larger than $10^5$.
The steady-state rotation speed is determined by the balance between the optical torque $M_z$ exerted on the nanodumbbell and the drag torque $T_z$ from air molecules. Since drag torque is proportional to the air pressure, the steady state rotation speed $\Omega_{rot}$ is inversely proportional to the air pressure $p_{air}$.  Fig. \ref{fig:rotation}(c) shows the rotation speed as a function of the air pressure. The experimental data agrees with $A/p_{air}$, where $A$ is a fitting parameter. As the air pressure decreases, the rotation speed increases. The maximum rotation frequency that we can measure is limited by the bandwidth of our photodetector.

As the rotation speed increases, eventually the nanodumbbell will fall apart due to the centrifugal force.
The ultimate rotating frequency  is determined by the ultimate tensile strength (UTS) of the material. Thus this experiment provides an opportunity to study material properties under extreme conditions.
The stress distribution of a nanodumbbell rotating at high speed can be calculated by the finite element method.
Figure \ref{fig:rotation}(d) shows the stress distribution of a $D=170$ nm, $L/D=1.9$ silica nanodumbbell rotating at 1.2 GHz. We assume the contacting point has a curvature radius of 5 nm. Remarkably, the maximum stress of the nanodumbbell under these conditions is about 13 GPa, which is 2 orders larger than the UTS of a bulk glass. This shows that our silica nanodumbbells   is as strong as state-of-the-art silica nanowires \cite{Brambilla2009ultimate}. The finite element method can be used to calculate the ultimate rotating frequency of silica nanodumbbells with different diameters and aspect ratios.
The range of ultimate rotating frequency is calculated by setting the UTS to be in the range of 10-20 GPa, which agrees with our observation and the results for silica nanowires\cite{Brambilla2009ultimate}.
Figure \ref{fig:rotation}(e) shows the calculated ultimate frequency is in the range of $1.1-1.6$ GHz  for $D=170$ nm, $L/D=1.9$ nanodumbbells. The ultimate frequency increases as the size of the nanodumbbell decreases.
Figure \ref{fig:rotation}(f) shows that for a given diameter $D=100$ nm, the ultimate rotation frequency increases when the aspect ratio decreases. Thus nanodumbbells with smaller diameters and smaller aspect ratios can sustain rotation frequencies beyond 10 GHz.

In conclusion, we have synthesized and optically trapped nanodumbbells at pressures below $10^{-4}$ torr without feedback cooling. With a circularly polarized laser, 170 nm-diameter nanodumbbells were driven to rotate beyond 1 GHz, which is the highest reported rotation frequency for a nanoparticle. Such high speed rotation may be used to study material properties and vacuum friction \cite{schuck2018ultrafast,nagornykh2017optical,Zhao2012rotational,Guo2014giant}. The torsional vibration of the levitated nanodumbbells in high vacuum is observed  when the laser is linearly polarized, an important step towards detecting the Casimir torque \cite{Xu2017detecting} and developing a quantum Cavendish torsion balance for studying the quantum nature of gravity \cite{Carlesso2018when,Marletto2017Gravitationally,Bose2017spin}. Multiple nanodumbbells can be levitated  together to study nonequilibrium phases and self-assembly \cite{lechner2013cavity}.






{\em Note added.} After completing this work, we became aware of a related work on fast rotation of a nanoparticle \cite{Reimann2018}.


%

\end{document}